\RequirePackage{ifpdf}

\documentclass[aps,prl,reprint,nofootinbib]{revtex4-1}

\usepackage{aj-definitions1}

\begin{document}

\title{Equilibrium partition function for nonrelativistic fluids}
\author{Nabamita Banerjee$^1$} 
\email[]{nabamita@iiserpune.ac.in}
\author{Suvankar Dutta$^2$} 
\email[]{suvankar@iiserb.ac.in}
\author{Akash Jain$^3$}
\email[]{akash.jain@durham.ac.uk}
\affiliation{$^1$Dept. of Physics, Indian Institute of Science Education
  and Research, Pune, $^2$Dept. of Physics, Indian Institute of Science Education
  and Research, Bhopal, $^3$Dept. of Mathematical Sciences \& Centre for Particle Theory, Durham University.}


\begin{abstract}
  We construct an equilibrium partition function for a non-relativistic fluid and use it to
  constrain the dynamics of the system. The construction is based on light cone reduction, which is
  known to reduce the Poincar\'e symmetry to Galilean in one lower dimension. We modify the
  constitutive relations of a relativistic fluid, and find that its symmetry broken phase -- `null
  fluid' is equivalent to the non-relativistic fluid. In particular, their symmetries,
  thermodynamics, constitutive relations, and equilibrium partition function match exactly to all
  orders in derivative expansion.
\end{abstract}


\maketitle



The constitutive relations of a relativistic fluid at local thermal equilibrium can be obtained from
an equilibrium partition function up to some undetermined `transport coefficients'
\cite{Banerjee:2012iz,Jensen:2012jh}. These coefficients can be determined either from experiments
or through a microscopic calculation. Recently, non-relativistic geometry and fluid are getting
active attention \cite{Son:2008ye, Son:2013rqa, Jensen:2014aia, Jensen:2014ama, Jensen:2014wha,
  Hartong:2014pma, Geracie:2015xfa}. If we think of non-rel fluid as a limit of an underlying
relativistic theory, we would expect its constitutive relations also to follow from an equilibrium
partition function. \emph{Goal of this letter is to develop a formal and consistent way to compute
  partition function for a non-relativistic fluid starting from a relativistic theory.}

We start with a flat background $\df s^2 = -2 \df x^- \df x^+ + \sum_{i=1}^d (\df x^{i})^2$, which
has $(d+2)$-dim Poincar\'e invariance. It is known that $(d+1)$-dim Galilean algebra sits inside
Poincar\'e -- all transformations that commute with $P_- = \dow_-$ (c.f. \cite{Duval:1984cj}). Hence
a theory on this background which respects $x^-$ independent isometries
$x^\sM \ra x^\sM + \xi^\sM(x^+,\vec x)$, $x^\sM = \{x^-,x^+,x^i\}$, enjoys Galilean
invariance. Compactifying the $x^-$ direction, we can realize this Galilean invariance as
non-relativistic invariance in $(d+1)$-dim \footnote{In this work we use `non-relativistic' as an
  alias for `Galilean', while the two are known to have some technical differences.}. This procedure
to get non-relativistic theories is known as \emph{light cone reduction (LCR)}.

Turning on $x^-$ independent fluctuations around the flat background,
\bem{\label{E:first-metric}
\df s^2 = \rmG_{\sM\sN} \df x^\sM \df x^\sN = g_{ij}\df x^i\df x^j\\
-2 e^{-\Phi} (\df x^+ + a_i \df x^i) (\df x^- - \cB_+ \df x^+ -\cB_i \df x^i) ,
}
non-relativistic theories can be described by a generating functional
$\cZ[\cB_+,\cB_i,\Phi,a_i,g_{ij}]$ written as a functional of background sources. Mapping Ward
identities of this partition function to those of a Galilean theory, allows us to read out the
Galilean currents and densities,
\bee{\label{E:expvar}
	\r \sim \frac{\d W}{\d \cB_+}, \
	j_{\r}^i \sim \frac{\d W}{\d \cB_i}, \
	\e \sim \frac{\d W}{\d \Phi}, \
	j^i_{\e} \sim \frac{\d W}{\d a_i}, \
	t^{ij} \sim \frac{\d W}{\d g_{ij}}, 
}
where $W = \ln\cZ$, and $\rho$, $j_{\r}^i$, $\e$, $j^i_{\e}$ and $t^{ij}$ are mass density, mass
current, energy density, energy current and stress tensor respectively of the non-rel theory.

In an earlier work \cite{Banerjee:2014mka}, following \cite{Rangamani:2008gi}, we performed LCR of a
relativistic fluid to obtain constitutive relations of a non-rel fluid (at leading order in
derivatives). While the reduced conservation equations agreed with their expected non-rel form (in
parity even sector), we observed that the non-rel fluid gained by reduction is not the most generic
one. In particular, thermodynamics that the reduced fluid follows is in some sense more restrictive
than the most generic non-rel fluid (chemical potential corresponding to particle number is not
independent). We also found that the parity-odd sector of reduced fluid survives only for a special
case of incompressible fluids. It strongly hints that to get the most generic non-rel fluid via
light cone reduction, we need to start with a modified relativistic system.

We start with a relativistic `fluid' on a curved background which admits a null isometry
\cite{Duval:1984cj,Julia:1994bs,Hassaine:1999hn,Christensen:2013rfa}. Unlike the `usual' relativistic fluids, now
isometry is also a background field and hence must be considered while writing the respective
constitutive relations. We break the relativistic symmetry to Galilean by only considering
diffeomorphisms which do not alter the isometry, and term the symmetry-broken relativistic fluid as
`null fluid'. This simple consideration happens to resolve all the issues we enlisted before. In
fact it does much more that that; even before LCR, $(d+2)$-dim null fluid is essentially equivalent
to a $(d+1)$-dim non-rel fluid, as they have same symmetries. As we shall show, their constitutive
relations, conservation equations, thermodynamics etc. match exactly to all orders in derivative
expansion. In this letter we aim to use this equivalence to write an eqb. partition function for
Galilean fluids, and use \cref{E:expvar} to constrain the dynamics of the system.

\vspace{.3cm}

\noindent {\bf Null Backgrounds:} 
We start with a $(d+2)$ dimensional spacetime $\cM_{(d+2)}$ equipped with a metric $\rmG_{\sM\sN}$ and
a metric compatible affine connection $\hat\G^\sR_{\ \sM\sN}$ (with associated covariant derivative
$\hat\N_{\sM}$). On this setup we define a null isometry $V$ ($\Ra\cL_{V}G_{\sM\sN} = 0$) which
satisfies $\hat\N_\sM V^\sN = 0$. We call this background `\emph{null background}'
\footnote{Standard terminology for `null backgrounds' is `Bargmann structures'. We use the former to
  be consistent with the charged case discussed in \cite{charged_nullfluid} where the two
  definitions are different.}. On torsion-less null backgrounds, which we shall consider in this
work, the latter condition implies the former, and in addition:
$\cH_{\sM\sN} = 2\partial_{[\sM} V_{\sN]}= 0$.  This is a dynamic constraint on the background, and can be
violated by quantum fluctuations off-shell.

A physical theory living on a null background, can be characterized by ($\log$ of) a generating functional $W$,
whose response to infinitesimal variation of metric is given by,
\bee{\label{E:RELPF} \d W = \int \lbr \df x^\sM \rbr \sqrt{-\rmG} \ \half T^{\sM\sN} \d
  \rmG_{\sM\sN}.  }
$T^{\sM\sN}$ is called the energy-momentum tensor. One can check that $V$ being null allows for an
arbitrary redefinition $T^{\sM\sN} \ra T^{\sM\sN} + \q V^\sM V^\sN$ for some scalar $\q$, which leaves \cref{E:RELPF}
invariant.  Demanding this partition function to be invariant under $V$ preserving diffeomorphisms,
we get the conservation law for energy-momentum, 
\begin{equation} \label{E:relcons}
  \hat\N_\sM T^{\sM\sN} = 0.
\end{equation}
A background is said to be in equilibrium configuration if it admits a timelike Killing vector
$K^\sM$. We right away choose a basis $x^\sM = \{x^-,x^+,x^i\}$, such that $V = \dow_-$ and
$K = \dow_+$. 
The most generic metric with this choice of basis is given by \cref{E:first-metric}.  It is easy to
verify that under $(d+2)$-diffeomorphisms restricted to our choice of basis, $\F, \cB_+$ transform
as scalars, $a_i, B_i(= \cB_i - a_i \cB_+)$ transform as abelian vector gauge fields, and $g_{ij}$
transforms as a metric in $d$-dim. Hence at equilibrium, the partition function $W^{eqb}$ will be a gauge
invariant scalar made out of these fundamental fields. Its variation can be worked out trivially
from \cref{E:RELPF},
\bem{\label{E:deqbreducedPF}
	\d W^{eqb} = \int \lbr \df x^i \rbr \sqrt{g} \frac{1}{\vq_o} \lB 
		\E{\F}\lb T_{+-} + T_{--} \cB_+ \rb \frac{1}{\vq_o} \d \vq_o \dbrk
		+ T^i_{\ +} \d a_i 
		+ \frac{1}{2} T^{ij} \d g_{ij} 
		+ \vq_o T_{--} \d \vp_o - T^{i}_{\ -} \d B_i
	\rB.
}
In getting this we have identified inverse temperature of the euclidean field theory to be $1/\tilde
\vq = \tilde\b\tilde R$, where $\tilde\b$ is the radius of the euclidean time circle, and
$\tilde R$ is the radius of the compactified null direction. Further we have defined the redshifted
equilibrium temperature $\vq_o = \tilde \vq \E{\F}$ and mass chemical potential $\vq_o\vp_o = \cB_+ \E{\F}$.

$W^{eqb}$ can be written order by order in derivatives of the source fields. While the partition
function is itself gauge and diffeo invariant, it is not true for the integrand. In fact,
\emph{Chern-Simons} terms can be added to it whose variation is gauge invariant only upto some boundary
terms. We shall consider such terms when required. At ideal order, $W^{eqb}$ is given by a function
of $\vq_o, \vp_o$,
\bee{ W_{o}^{eqb} = \int \lbr \df x^i \rbr \sqrt{g}
  \frac{1}{\vq_o} P_o (\vq_o,\vp_o), \label{nulPF_ideal}}
where $P_o$ is the local thermodynamic \emph{pressure} at equilibrium. Varying it and using
\cref{E:deqbreducedPF}, we will get,
\bee{\label{E:ideal_PFrel}
	T^{ij} = P_o g^{ij}, \ \
	T_{--} = R_o, \ \
	\E{\F}(T_{+-} + T_{--}\cB_o) = E_o,
}
where considering $\vq_o,\vp_o$ as thermodynamic variables at equilibrium, we have defined the first
law of thermodynamics and Gibbs-Duhem equation,
\bee{\label{E:thermo}
	\df E = \vq \df S + \vq \vp \df R, \quad E =  S\vq + \vq R \vp - P.
}
Defining a null field $\bar V_{(K)}^\sM = \E{\F} \lb K^\sM + \cB_+ V^\sM \rb$ normal to $V$
(i.e. $V^\sM \bar V_{(K)\sM} = -1$), and a projection operator $P^{\sM\sN}_{(K)} = \rmG^{\sM\sN} + 2
V^{(\sM}\bar V^{\sN)}_{(K)}$ transverse to $V$, $\bar V_{(K)}$, \cref{E:ideal_PFrel} can be covariantly
repackaged into,
\bee{\label{E:RelCons_ideal}
	T^{\sM\sN} = R_o \bar V_{(K)}^\sM \bar V_{(K)}^\sN + 2 E_o \bar V_{(K)}^{(\sM} V^{\sN)} + P_o P_{(K)}^{\sM\sN}.
}
\noindent {\bf Null Fluid:} 
Having constructed null backgrounds, we proceed to define hydrodynamics on this setup. This is the
essence of our work; we claim that this `modified fluid' is just a different representation of the
non-rel fluid, with exact (yet trivial) mapping between the two facilitated by light cone reduction. Note that conservation laws (\ref{E:relcons}) are $(d+2)$ equations, so any system with
$(d+2)$ independent variables would be exactly solvable on this background. We choose to describe
our system by a fluid, with a null velocity field $u^\sM$ normalized as
$u^\sM u_\sM = 0, u^\sM V_\sM = -1$ and two thermodynamic variables, temperature $\vq$, and mass
chemical potential $\vp$.  The most generic
constitutive relations (after using the $T^{\sM\sN}$ redefinition freedom) are given in terms of fluid
variables and background quantities as,
\bem{\label{E:RelCons}
	T^{\sM\sN} = \cR u^\sM u^\sN + 2 \cE u^{(\sM}V^{\sN)} + \cP P_{(u)}^{\sM\sN} +
        2 \bbR^{(\sM}u^{\sN)} \\
        + 2 \bbE^{(\sM}V^{\sN)} + \bbT^{\sM\sN},
}
where $\cR,\cE,\cP$ are some arbitrary functions of $\vq,\vp$.  The tensors
$\bbR^\sM, \bbE^\sM, \bbT^{\sM\sN}$(traceless) contain derivative corrections and are transverse to
$u^\sM$ and $V^\sM$ through projection operator: $P_{(u)}^{\sM\sN} = \rmG^{\sM\sN} + 2V^{(\sM}u^{\sN)}$.

From \cref{E:RelCons_ideal} we can deduce that at equilibrium (ideal order), $\cR,\cE,\cP$ and $u^\sM$ boil down to
the thermodynamic functions $R,E,P$ and equilibrium null vector $\bar V_{(K)}^\sM$. Outside
equilibrium however, none of the fluid variables are uniquely defined, and are subjected to
arbitrary redefinitions. 
These are two scalars and a vector worth of freedom, which we fix by working in `mass frame',
i.e. we identify $\cE,\cR$ with $E,R$ dumping all the dissipation into $\cP$, and set $\bbR^\sM = 0$,
\bem{\label{E:MassFrame}
	T^{\sM\sN} = R u^\sM u^\sN + 2 E u^{(\sM}V^{\sN)} + P P_{(u)}^{\sM\sN} + 2
        \cE^{(\sM}V^{\sN)} \\ 
	+ \Pi^{\sM\sN}.
}
Here $\Pi^{\sM\sN}$ is not traceless. To leading derivative order (one derivative in parity-even sector and $(n-1)$ derivative in parity-odd sector for $d = 2n-1$), constitutive relations in mass frame are given as:
\bea{\label{E:rel1oddeven_CR}
	\Pi^{\sM\sN} &= - \eta \s^{\sM\sN} - P_{(u)}^{\sM\sN}\z \Q, \nn\\
	\cE^{\sM} &= P_{(u)}^{\sM\sN} \lB \k  \dow_\sN \vq + \l \dow_\sN \vp \rB 
	+ \tilde\o l^\sM,
}
where $\s^{\sM\sN} = 2 P_{(u)}^{\langle \sM\sR}P_{(u)}^{\sN\rangle \sS} \hat\N_{\sR} u_\sS$ is symmetric traceless, $\Q = \hat\N_\sM u^\sM$ and $l^\sM = \star [ V \wedge u \wedge (\df u)^{\wedge (n-1)} ]^\sM$. \\


\noindent{\it Equilibrium Partition Fuction}: Similar to relativistic fluids in \cite{Banerjee:2012iz,Jensen:2012jh}, equilibrium partition function 
gives equality type constraints among various transport coefficients appearing in the null fluid constitutive relations. Away from ideal order we can construct the partition function $W^{eqb}$ order by order in derivatives of
the background fields. It is easy to see that at leading
derivative order, there are no scalars at equilibrium and the partition function is trivially zero
upto some Chern-Simons terms: 
\bee{\label{E:1PF}
	W^{eqb}
	= W_{o}^{eqb} - \int \lb n C_1 \tilde \vq a + C_2 B \rb \wedge (\df B)^{\wedge (n-1)}.
}
The term coupling to $C_1$ goes as $B\wedge \df a \wedge (\df B)^{\wedge (n-2)}$ upto a total derivative term, which vanishes on-shell as $\cH_{ij} = 0 \Ra \df a = 0$. But since partition functions are to be written off-shell, we must include this term. On the other hand, in constitutive relations \cref{E:rel1oddeven_CR}, only terms coupling to $\l$ and $\tilde\o$ survive at equilibrium. Comparing these to the constitutive relations generated by partition function variation defined in \cref{E:deqbreducedPF}, one can easily get the constraints:
\bee{\label{E:constraints}
	\l = 0, \qquad
	\tilde\o =
	n \vq  \lb
		\vq C_1
		+ \frac{E + P - \vq\vp R}{R} C_2
	\rb.
}
Fluid variables $\vq,\vp$ do not get corrections at leading order, while velocity gets a parity-odd correction: 
\bee{
	u^{\sM}
	= \bar V_{(K)}^\sM - \frac{\vq_o n}{R_o} C_2 \star [(\df B)^{\wedge (n-1)}]^\sM.
} 

\noindent{\it Entropy Current:}
Second law of thermodynamics demands that there must exist an entropy current $J^\sM_s$, whose
divergence is positive semi-definite $\hat\N_\sM J^\sM_s \geq 0$. The most generic form of entropy
current is given as, $J_{s}^\sM = J_{s(can)}^\sM + \U^\sM_s$, where $\U^M_s$ are arbitrary derivative
corrections (not necessarily projected), and,
\bee{\label{E:EC}
	J_{s(can)}^\sM = \frac{1}{\vq} P u^\sM - \frac{1}{\vq} T^{\sM\sN} u_\sN + \vp T^{\sM\sN} V_\sN,
}
is the canonical entropy current. Its divergence can be computed to be,
\bee{
	\vq\hat\N_\sM J_{s(can)}^\sM 
	=
	- \Pi^{\sM\sN} \hat\N_\sM u_\sN
	- \frac{1}{\vq} \cE^{\sM} \hat\N_\sM \vq.
}
Plugging in the constitutive relations \cref{E:rel1oddeven_CR} and only allowed derivative
correction $\U^\sM_s = \tilde\o_s l^\sM$ (other vectors give pure derivative terms in divergence and
hence must vanish), one can find that second law of thermodynamics gives the same constraints
\cref{E:constraints} (in parity-even sector) and in addition: $\eta, \z \geq 0$ and $\k \leq 0$. In
parity-odd sector however, it sets $C_2 = 0$ and $C_1 = C_1 (\vq)$ which unexpectedly is weaker than
the partition function constraints. This discrepancy can be accounted to the fact that in this
computation we have missed constraint(s) coupling to $\cH = \df V$, which is set to zero by
torsionlessness requirement. This condition can however be violated off-shell and hence the coupled
constraint(s) are visible to equilibrium partition function. In a companion paper
\cite{charged_nullfluid}, we will show that on introducing just enough torsion to allow for non-zero
values of $\cH$, and setting it to zero after the entropy current computation, we will recover the
missed constraints.

\vspace{.3cm}

\noindent {\bf Light Cone Reduction:} To give null backgrounds a Galilean interpretation, we need to
get rid of the isometry direction $V$. To do so, we curl up the $V$ direction into an infinitesimal
circle, reducing the effective dimensions by one, where non-relativistic theory can live. To make this more
precise, we pick up an arbitrary vector field $T (\neq V)$ on $\cM_{(d+2)}$, and use it to define a unique foliation
of $\cM_{(d+2)} = S^1_V \times R^1_T \times \cM_{(d)}^T$, where,
\bee{
\cM_{(d)}^{T} := \lbr v^\sM: v^\sM V_\sM = v^\sM T_\sM = 0 \rbr.
}
%
%
After this point one just has to choose a basis on this foliation and read out the Galilean results,
i.e. $x^\sM = \{ x^-, x^+, x^i \}$ such that $V = \dow_-$, $T = \dow_+$, and $\vec x = \{x^i\}$ span
the spatial manifold $\cM_{(d)}^T$. Note that this basis depends on the choice of `Galilean frame'
$T$, which we will refer to as \emph{local rest} of a frame $T$. One could also work in a frame
independent \emph{Newton-Cartan} formalism, which we discuss in a followup paper
\cite{charged_nullfluid}.

One can check that with this choice of basis, metric $\rmG_{\sM\sN}$ decomposes as
\cref{E:first-metric}, with all its components being independent of $x^-$.  Using this
decomposition, partition function variation \cref{E:RELPF} can be reduced to,
\bem{\label{E:dKKreducedPF}
	\d W = \tilde R \int \df x^+ \lbr \df x^i \rbr \sqrt{g} \ \E{-\F} \Big[
		- (
			\E{-\F} j_\e^i 
			- j_\r^i \cB_+
		) \d a_i \\
		+ \hat \e_{tot} \d \F
		+ \half t^{ij} \d g_{ij} 
		+ \lb \E{\F} \hat \r \d \cB_+ + j_\r^i \d B_i \rb 
	\Big],
}
where $\tilde R$ is the radius of compactified $x^-$, and we have identified,
\bea{\label{identifi}
	\hat \r &= T_{--}, \quad
	\hat \e_{tot} = \E{\F} \lb T_{-+} + T_{--}\cB_+ \rb, \quad
	t^{ij} = T^{ij} \nn \\
	j_\r^i &= - T^{i}_{\ -} , \quad
	j_\e^i = - \E{\F} \lb T^i_{\ +} + T^i_{\ -}\cB_+ \rb.
}
Using these identifications, we can reduce the Ward identities \cref{E:relcons} to get,
\bea{\label{E:inertial_CR}
	\frac{1}{\sqrt{g}}\dow_+ \lb \sqrt{g} \r\rb + \N_i \lb \E{-\F}  j_\r^i \rb &= 0, \nn \\
	\frac{1}{\sqrt{g}}\dow_+ \lb \sqrt{g} \e_{tot}\rb + \N_i \lb \E{-\F}  j_\e^i \rb
	&= - \half t^{ij} \dow_+ g_{ij} - \E{-\F} j_\r^i \a_i, \nn \\
	\frac{1}{\sqrt{g}}\dow_+ \lb\sqrt{g} p_i \rb
	+ \N_j \lb \E{-\F} t^{j}_{\ i} \rb
	&= - \half \E{-\F} a_i t^{jk} \dow_+ g_{jk}, \nn\\
        - \E{-\F} &\lb \hat \r \E{-\F} \a_i + j^j_{\r}  \o_{ji} \rb,
}
where we have defined corrected densities due to time not being `flat',
$\r = \hat \r - \E{-\F} j_\r^i a_i$, $\e_{tot} = \hat \e_{tot} - \E{-\F} j_\e^i a_i$, and
$p^i = j_\r^i - \E{-\F} t^{ij} a_j$. Identifying $x^+$ with the Galilean time, these equations can
be realized as mass, energy and momentum conservation laws respectively of a Galilean theory. Mass
is exactly conserved, while energy/momentum are being sourced due to time-dependence of background
metric.  Further, energy/momentum are also being sourced due to pseudo-energy/force caused by
acceleration $ \a_i = - \E{\F} [\df \cB]_{i+}$ and vorticity $\o^{ij} = [\df \cB]^{ij}$ of the
Galilean frame.

\vspace{.3cm}



\noindent{\bf Non-Relativistic Fluid:} It is interesting to see how the dynamics of a non-relativistic fluid 
emerges from the aforementioned reduction. The conservation equations \cref{E:inertial_CR} are in one-to-one
correspondence with the dynamical equations of a non-relativistic fluid as given in \cite{Banerjee:2014mka},
generalized to curved space. This motivates us to interpret null fluid on $\cM_{(d+2)}$ as a
Galilean fluid on $\bbR_T^1 \times \cM_{(d)}^T$, as seen by some reference frame $T$. We can right
away use \cref{E:RelCons} as an ansatz for $T^{\sM\sN}$ and perform the reduction as suggested by \cref{identifi},
\bea{\nn
	\hat\e_{tot} &= \hat\e + \half \hat\r v^i v_i + \vs_\r^i v_i, \qquad
	j_\r^i = \hat\r v^i + \vs_\r^i, \nn\\
	t^{ij} &= P g^{ij} + \hat\r v^i v^j + 2 v^{(i} \vs_\r^{j)} + \pi^{ij}, \nn\\
	j_\e^i &= \lb \hat\e_{tot} + P \rb v^i + \vs_\e^i + \pi^{ik} v_k + \half \vs_\r^i v^j v_j,
}
with identifications:
\bee{\nn
	\hat\r = \cR, \
	\hat\e = \cE, \
	\vs_\r^i = \bbR^i, \
	\vs_\e^i =  \bbE^{i}, \
	\pi^{ij} = (\cP - P) g^{ij} + \bbT^{ij}.
}
Dynamics of these densities/currents is governed by conservation laws \cref{E:inertial_CR}. Entropy
current \cref{E:EC} on the other hand reduces to:
\bee{\label{E:EC}
	\hat s = \frac{\hat\e + P}{\vq} - \vp \hat\r + \U_{s-}, \quad
	j_{s}^i = \hat s v^i + \frac{1}{\vq} \vs_\e^i - \vp \vs_\r^i + \U^i_s.
}
$\hat s = S$ at ideal order in equilibrium. The statement of second law of thermodynamics then becomes $\dow_+ \lb \sqrt{g} s\rb + \dow_i \lb \sqrt{g}\E{-\F}  j_s^i \rb \geq 0$, where $s = \hat s - \E{-\F}  j_s^i a_i$. 

As an example we can consider leading order fluid on flat background in three spatial dimensions ($d=3$) expressed in mass frame ($\vs_\r^i = 0$), as given in \cite{landau},
\bea{
	j_\r^i &= R v^i, \qquad \hat\e_{tot} = E + \half R v^i v_i, \nn\\
	t^{ij} &= R v^i v^j + P g^{ij} - 2\eta \dow^{\langle i} v^{j\rangle} - \z g^{ij} \dow_k v^k, \nn\\	
	j_\e^i &= \lb E + P + \half R v^i v_i \rb v^i - 2\eta v_j \dow^{\langle i} v^{j\rangle} - \z v^i \dow_k v^k \nn\\
	& \qquad + \k\dow^i \vq + \l\dow^i \vp + \tilde\o \e^{ijk}\dow_{j} v_k, \nn \\
	j_{s}^i &= S v^i + \frac{1}{\vq} \lB\k\dow^i \vq + \l\dow^i \vp + (\tilde\o + \vq \tilde\o_s) \e^{ijk}\dow_{j} v_k \rB
}
Demanding second law of thermodynamics to hold will give same constraints as the null fluid; in
particular $\l = 0$. As expected, parity-even sector contains a bulk viscosity, a shear viscosity
and a thermal conductivity term. Parity-odd sector however has a thermal Hall conductivity term
coupled to fluid vorticity. These constitutive relations follow the conservation laws
\cref{E:inertial_CR} restricted to flat space. Here we explicitly chose to work in mass frame, any
other choice of frame in the null fluid will follow trivially to the non-relativistic fluid due to
trivial mapping of currents.


\vspace{.2cm}

\noindent
{\it Equilibrium partition function:} From the perspective of a non-rel fluid, equilibrium is
defined by a preferred reference frame $K$ with respect to which system does not evolve in time. The
variation of eqb. partition function in local rest of reference frame $K$ is essentially same as the
null fluid \cref{E:deqbreducedPF} written in terms of Galilean quantities, and hence,
\bea{\nn
	\hat\r_o &= \frac{\d W^{eqb}}{\d \vp_o}, \quad
	j_{o\r}^i = \vq_o\frac{\d W^{eqb}}{\d B_i}, \quad
	t_o^{ij} = 2\vq_o\frac{\d W^{eqb}}{\d g_{ij}}, \nn\\
	\hat\e_{o} &= \vq_o^2\frac{\d W^{eqb}}{\d \vq_o}, \quad
	j^i_{o\e} - \vp_o\vq_o j_{o\r}^i  = - \vq^2_o\E{\F}\frac{\d W^{eqb}}{\d a_i}.
}
These will reduce to the expected relations \cref{E:expvar} in flat space $\vq_o = 1$, $g_{ij} =
\d_{ij}$, $\vp_o = B_i = a_i = 0$. In equilibrium, null-fluid and Galilean fluid have same field
content and symmetries, so we expect the eqb. partition function to also be the same,
i.e. \cref{nulPF_ideal,E:1PF}. To ideal order it will identify $\hat\r_o, \hat\e_o$ with the
thermodynamic functions $R,E$, and hence will give physical interpretation to the thermodynamics of
null fluids \cref{E:thermo} in terms of non-rel physics. At leading derivative order it will give
constraints \cref{E:constraints}. 

We would like to note here that \cite{Jensen:2014ama} also constructed an eqb. partition
function and entropy current for an uncharged Galilean fluid upto leading order in derivatives, but
purely from a Galilean perspective, without invoking null reduction or a relativistic
system. As expected, we find our constitutive relations and eqb. partition function to be in exact
agreement with \cite{Jensen:2014ama}.

\vspace{.3cm}

\noindent {\bf Discussion:} 
One of the most striking features of our construction is that the relativistic null fluid is
equivalent to a Galilean fluid, and is related just by a choice of basis $V = \dow_-$. This gives us
a new and rather simplified way to look at Galilean fluids altogether, since we have all the
machinery of relativistic hydrodynamics at our disposal. In the current work we have used it to
construct an eqb. partition function and an entropy current for torsionless Galilean fluids at
leading order in derivatives, and to find constraints on various transport coefficients appearing in the
fluid constitutive relations. The procedure can also be extended to include an (anomalous) $\rmU(1)$
current which we consider in a companion paper \cite{charged_nullfluid}. In another paper
\cite{akash} we use this idea of null backgrounds to study (non-abelian) gauge and gravitational
anomalies in (torsional) Galilean theories, and in particular their effect on hydrodynamics.

\vspace{.3cm}


\noindent {\bf Acknowledgments:} We are thankful to Arjun Bagchi, Ashish Kakkar and Parikshit Dutta for useful discussions. The work of NB is supported by DST Ramanujan Fellowship. AJ would like to thank Durham Doctoral Scholarship for financial support.


\begin{thebibliography}{9}


Literature on non-relativistic hydrodynamics is vast and is increasing every day, thus it is not possible to give an exhaustive list of references. We plan to include a more complete list of references in a follow-up work.

\bibitem{Banerjee:2012iz} 
  N.~Banerjee, J.~Bhattacharya, S.~Bhattacharyya, S.~Jain, S.~Minwalla and T.~Sharma,
  ``Constraints on Fluid Dynamics from Equilibrium Partition Functions,''
  JHEP {\bf 1209}, 046 (2012)
  [arXiv:1203.3544 [hep-th]].

\bibitem{Jensen:2012jh} 
  K.~Jensen, M.~Kaminski, P.~Kovtun, R.~Meyer, A.~Ritz and A.~Yarom,
  ``Towards hydrodynamics without an entropy current,''
  Phys.\ Rev.\ Lett.\  {\bf 109}, 101601 (2012)
  [arXiv:1203.3556 [hep-th]].

\bibitem{Son:2008ye} 
  D.~T.~Son,
  ``Toward an AdS/cold atoms correspondence: A Geometric realization of the Schrodinger symmetry,''
  Phys.\ Rev.\ D {\bf 78}, 046003 (2008)
  [arXiv:0804.3972 [hep-th]].

\bibitem{Son:2013rqa} 
  D.~T.~Son,
  ``Newton-Cartan Geometry and the Quantum Hall Effect,''
  arXiv:1306.0638 [cond-mat.mes-hall].

\bibitem{Jensen:2014aia} 
  K.~Jensen,
  ``On the coupling of Galilean-invariant field theories to curved spacetime,''
  arXiv:1408.6855 [hep-th].
  
\bibitem{Jensen:2014ama} 
  K.~Jensen,
  ``Aspects of hot Galilean field theory,''
  JHEP {\bf 1504}, 123 (2015)
  [arXiv:1411.7024 [hep-th]].
  
\bibitem{Jensen:2014wha} 
  K.~Jensen and A.~Karch,
  ``Revisiting non-relativistic limits,''
  JHEP {\bf 1504}, 155 (2015)
  [arXiv:1412.2738 [hep-th]].

\bibitem{Hartong:2014pma} 
  J.~Hartong, E.~Kiritsis and N.~A.~Obers,
  ``Schroedinger Invariance from Lifshitz Isometries in Holography and Field Theory,''
  arXiv:1409.1522 [hep-th].

\bibitem{Geracie:2015xfa} 
  M.~Geracie, K.~Prabhu and M.~M.~Roberts,
  ``Fields and fluids on curved non-relativistic spacetimes,''
  arXiv:1503.02680 [hep-th].

\bibitem{Duval:1984cj} 
  C.~Duval, G.~Burdet, H.~P.~Kunzle and M.~Perrin,
  ``Bargmann Structures and Newton-cartan Theory,''
  Phys.\ Rev.\ D {\bf 31}, 1841 (1985).

\bibitem{Banerjee:2014mka}
  N.~Banerjee, S.~Dutta, A.~Jain and D.~Roychowdhury,
  ``Entropy current for non-relativistic fluid,''
  JHEP {\bf 1408} (2014) 037
  [arXiv:1405.5687 [hep-th]].

\bibitem{Rangamani:2008gi}
  M.~Rangamani, S.~F.~Ross, D.~T.~Son and E.~G.~Thompson,
  ``Conformal non-relativistic hydrodynamics from gravity,''
  JHEP {\bf 0901}, 075 (2009)
  [arXiv:0811.2049 [hep-th]].

\bibitem{Julia:1994bs} 
  B.~Julia and H.~Nicolai,
  ``Null Killing vector dimensional reduction and Galilean geometrodynamics,''
  Nucl.\ Phys.\ B {\bf 439}, 291 (1995)
  [hep-th/9412002].

\bibitem{Hassaine:1999hn} 
  M.~Hassaine and P.~A.~Horvathy,
  ``Field dependent symmetries of a nonrelativistic fluid model,''
  Annals Phys.\  {\bf 282}, 218 (2000)
  [math-ph/9904022].

\bibitem{Christensen:2013rfa} 
  M.~H.~Christensen, J.~Hartong, N.~A.~Obers and B.~Rollier,
  ``Boundary Stress-Energy Tensor and Newton-Cartan Geometry in Lifshitz Holography,''
  JHEP {\bf 1401}, 057 (2014)
  [arXiv:1311.6471 [hep-th]].

\bibitem{landau}
  L.~D.~Landau \& E.~M.~Lifshitz (1997). Fluid mechanics, Pergamon Press.

\bibitem{charged_nullfluid} 
  N.~Banerjee, S.~Dutta and A.~Jain,
  ``Null Fluids -- A New Viewpoint of Galilean Fluids,''
  arXiv:1509.04718 [hep-th].

\bibitem{akash} 
  A.~Jain,
  ``Galilean Anomalies and Their Effect on Hydrodynamics,''
  arXiv:1509.05777 [hep-th].

  

\end{thebibliography}
\end{document}